\documentclass[aps,prb,twocolumn,superscriptaddress,showpacs,floatfix,amsmath,amssymb]{revtex4}
\usepackage{graphicx}
\usepackage{color}
\usepackage[normalem]{ulem}

\DeclareMathAlphabet\mathbfcal{OMS}{cmsy}{b}{n}

\newcommand{\br}[0]{ {\bf r} }
\newcommand{\iind}[4]{ \iint\!\!\mathrm{d}^{#1}#2\mathrm{d}^{#3}#4 \; }
\newcommand{\cW}[0]{ \hat{ W } }
\newcommand{\FP}[0]{ \hat{\Psi} }
\newcommand{\FPd}[0]{ \hat{\Psi}^\dagger }
\newcommand{\PO}[0]{ \Phi }
\newcommand{\POd}[0]{ \Phi^\dagger }

\newcommand{\FPupd}[0]{ \hat{\psi}^\dagger_{\uparrow} }

\newcommand{\FPdownd}[0]{ \hat{\psi}^\dagger_{\downarrow} }
\newcommand{\indll}[3]{ \int_{#2}^{#3}\!\!\mathrm{d}#1 \; }

\begin{document}

\author{S. Pittalis}
\email[]{stefano.pittalis@nano.cnr.it}
\affiliation{Istituto Nanoscienze, Consiglio Nazionale delle Ricerche, Via Campi 213A, I-41125 Modena, Italy}

\author{G. Vignale}
\affiliation{Department of Physics and Astronomy, University of Missouri, Columbia, Missouri 65211, USA}

\author{F. G. Eich}
\affiliation{Max Planck Institute for the Structure and Dynamics of Matter, Luruper Chaussee 149, D-22761 Hamburg, Germany}

\title{U(1)$\times$SU(2) Gauge Invariance Made Simple for Density Functional Approximations}

\date{\today}

\begin{abstract}
A semi-relativistic density-functional theory that includes spin-orbit couplings and Zeeman fields on equal footing with the electromagnetic potentials, is an appealing framework to develop a unified first-principles computational approach for non-collinear magnetism, spintronics, orbitronics, and topological states.  The basic variables of this theory include the paramagnetic current  and the spin-current density, besides the particle and the spin density, and the corresponding exchange-correlation (xc) energy functional is invariant under local  U(1)$\times$SU(2) gauge transformations. 
The xc-energy functional must be approximated to enable practical applications, but, contrary to the case of the standard density functional theory,   finding simple approximations suited to deal with realistic atomistic inhomogeneities has been a long-standing challenge. Here, we propose a way out of this impasse by showing that approximate gauge-invariant  functionals can be easily generated from existing  approximate functionals of ordinary density-functional theory by applying a simple {\it minimal substitution} on the kinetic energy density, which controls the  short-range behavior of the exchange hole.   Our proposal opens the way to the construction of approximate, yet non-empirical functionals, which do not assume weak  inhomogeneity and should therefore have a wide range of applicability in atomic, molecular and condensed matter physics. 
\end{abstract}


\maketitle

\section{Introduction} 
Density-functional methods are the most widely used approach to efficiently compute the electronic structure of atoms, molecules and solids. Based on the  Hohenberg-Kohn theorem~\cite{HohenbergKohn:64}, the electronic ground-state energy of interacting electrons is computed via the solution of the Schr{\"o}dinger equation for fictitious noninteracting electrons -- the so-called Kohn-Sham (KS) electrons~\cite{KohnSham:65}.  At the heart of Density-Functional Theory (DFT) lies the idea that the exchange-correlation (xc) energy, i.e., the energy due to the electron-electron interactions beyond the classical Hartree energy, can be approximated by a \emph{universal} functional of appropriate local densities. In its original incarnation, DFT considered only an external scalar potential coupled to the charge density, which characterizes the interacting system. This means that the universal functional for the xc energy could be written as a functional of the charge density alone. However, as soon as additional couplings, e.g., the Zeeman term or the coupling to an external vector potential, are present, the universality of the xc-energy functional is lost, unless additional densities are included as fundamental variables. This observation led over the years to the creation of {\it multivariate} DFTs, such as Spin-DFT (SDFT)~\cite{BarthHedin:72} for including the Zeeman coupling and Current-DFT (CDFT)~\cite{VignaleRasolt:87} for orbital magnetism.  In these theories, the spin density and the paramagnetic current density are included as basic variables, respectively.  For the description of two dimensional heterostructures~\cite{WinklerBook:03,Dresselhaus:55,RashbaSheka:59} or topological insulators~\cite{KaneMele05,LiangCheng11,Ando13}, Spin-Orbit Coupling (SOC) plays a crucial role~\cite{HasanKane:10}. Since SOC is naturally described in terms of spin-dependent vector potentials, its density-functional treatment requires the additional inclusion of spin current densities  as basic variables for a universal xc-energy functional. The corresponding extension of DFT has been dubbed Spin-Current-DFT (SCDFT)~\cite{VignaleRasolt:88,Bencheikh:03}.

Upgrading DFT to include additional variables not only leads to universal functionals for systems with SOCs and in strong magnetic fields,  
but also brings forth  an important physical concept, namely, gauge symmetry~\cite{FroehlichStuder:93}, which places strong constraints on the admissible dependence of the xc-energy functional on the basic variables and offers guidance in the construction of approximate functionals.  
In CDFT, for example,  the xc-energy functional is invariant under \emph{local} U(1) gauge transformations. 
Similarly, for spin-orbit coupled systems, the xc-energy functional is also invariant under \emph{local} SU(2) gauge transformations (we will come back to this point in the next section extensively). 
In order to guarantee the U(1) invariance of the theory, Vignale and Rasolt argued that the xc energy should not be expressed  in terms of the paramagnetic current, which is gauge dependent, but in terms of the gauge invariant vorticity~\cite{VignaleRasolt:87}: thus, they arrived at the first universal local density approximation for electrons in a magnetic field.   Subsequent experimentation showed that the U(1)-invariant vorticity is not well suited to deal with strongly inhomogeneous systems, such as atoms~\cite{ZhuTrickey:06}.  Later, Abedinpour, Vignale and Tokatly proposed the generalized U(1)$\times$SU(2) covariant vorticity as the fundamental variable for constructing gauge-invariant approximations to the xc-energy functional in SCDFT~\cite{AbedinpourTokatly:10}. Its definition depends on an {\em arbitrary}  choice of a ``linking path'' in physical space (cf.\ Ref.\ \onlinecite{AbedinpourTokatly:10} for details) -- an arbitrariness that, however, disappears in the limit of slowly varying densities.  
To the best of our knowledge, no experience has yet been  gained on the practical use of the SU(2)-covariant vorticity. In any case, this quantity
only emerges naturally when analyzing systems with {\em minor} inhomogeneities, such as an almost-uniform electron gas, and it may therefore suffer from the same shortcomings as its U(1) counterpart, when applied to atomistic systems. 

One lesson we learn from these examples is that the choice of  the gauge-invariant ``building blocks" of the DFT is a nontrivial and important task, which requires much ingenuity as well as extensive experimentation on realistic systems.  For example, the development of CDFT and, to a lesser extent, of SDFT,  has been hampered for many  years by the lack of functionals suitable to work with strong inhomogeneities.   In the case of CDFT, the difficulty arose from the vorticity being too closely tied  to the uniform free electron gas model.  In the case of SDFT, as applied to magnetic systems with non-collinear spins, it is well known that a straightforward generalization of functionals derived in the collinear SDFT framework~\cite{KueblerWilliams:88}  only accounts for longitudinal fluctuations of the spin magnetization. Various attempts  have been made recently~\cite{ScalmaniFrisch:12,EichGross:13,EichVignale:13} to  include a dependence on transverse fluctuation of the spin magnetization within SDFT, but none of them, for different reasons, has been proven fully satisfactory.

In this paper, we propose a way out of the impasse, following a suggestion by Tao and Perdew~\cite{TP05} 
who noticed that functionals of the Meta-Generalized-Gradient-Approximation (MGGA) family, such as the TPSS and similar~\cite{TPSS03}, which use the kinetic energy density $\tau({\bf r})$ as a basic variable,  
could be made current-dependent  by enforcing U(1) gauge invariance through the {\em minimal} substitution
\begin{equation}\label{TaoPerdew}
 \tau \rightarrow \tilde{\tau} =  \tau - \frac{{\bf j} \cdot {\bf j} }{ 2n }\,,
\end{equation}
where ${\bf j}$ is the {\em paramagnetic} current density.  
Earlier evidence of the relevance of 
their suggestion can be found in ideas by Dobson~\cite{Dobson93}  and, particularly,  Becke~\cite{Becke-j96}, whose current-dependent functional, based on a careful study of the short-range behavior of the exchange hole (x-hole),  greatly improved the description of degenerate ground states in open shell atoms~\cite{Becke-j02}.
Follow-up works extended these ideas and demonstrated their usefulness in applications~\cite{J-PBE, Pittalis-J, Tricky-J}.

Along similar lines, we show that DFT-MGGA forms can be readily upgraded to SCDFT-MGGA forms 
(i.e., they can be made spin-dependent and spin-current dependent) by enforcing the U(1)$\times$SU(2) gauge invariance through {\em minimal} substitutions to be performed on the curvature of the exchange hole.
These substitutions implicate new quantities such as the ``spin-kinetic energy" (defined below), the spin-currents {\em and} the spin density combined with  its gradients.
Additionally,  a trivial modification has to account for an extra dependence of the extended on-top exchange-hole 
on the squared modulus of the magnetization (and {\em no} other extra combinations).
In this manner, successful approximations of ordinary DFT can be readily turned into approximations for SCDFT and non-collinear SDFT.  
Our proposal opens the way to the construction of approximate, yet non-empirical functionals,  which do not assume weak  inhomogeneity 
and should therefore have a wide range of applicability in atomic, molecular, and condensed matter physics. 

This paper is organized as follows.  
In Sec.~\ref{Sec2}, we  review  the  theoretical background of U(1)$\times$ SU(2) gauge invariance in SCDFT and take the opportunity to remark that the xc-fields --  generated, as usual, through functional derivatives of the xc-energy functional with respect to the basic densities -- can exert non-trivial torques on the spin density and paramagnetic spin currents. 
In Sec~\ref{Sec3}, we derive the generalized short-range behavior of exchange-only quantities, which allows us to extract very useful
U(1) $\times$ SU(2) gauge-invariant quantities for the construction of  functional approximations in SCDFT and, therefore, we point out the
aforementioned  minimal substitutions. In Sec.~\ref{Sec4}, we give  examples of new approximate functionals of the MGGA form.
Technical details concerning the gauge transformations of the various fundamental quantities are reported in Appendix~\ref{Appendix}.

\section{Theoretical Background: Gauge Invariance and XC-torques}\label{Sec2} 

SCDFT is concerned with the calculation of the ground-state energy and densities of the semi-relativistic many-electron Pauli Hamiltonian~\cite{Landau3}.
In comparison to the SDFT Hamiltonian,  which contains only a scalar potential, $v({\bf r})$,  and a Zeeman magnetic field, $B^a({\bf r})$, the SCDFT also includes  an Abelian vector potential, ${\bf A}({\bf r})$, and a non-Abelian vector potential ${\bf A}^{a}({\bf r})$:~\cite{note1}
\begin{eqnarray}\label{H_1}
\hat{H} &=&  
\frac{1}{2} \int d^3r~ \hat{\Psi}^\dagger({\bf r}) \left[  -i \nabla + \frac{1}{c} {\bf A}({\bf r})  +  \frac{\mu_B}{2c}    \sigma^{a}  {\bf A}^{a}({\bf r}) \right]^2 
\hat{\Psi}({\bf r})  \nonumber \\
&+& \cW + \int d^3r~\hat{n}({\bf r}) v({\bf r}) + \mu_B \int d^3r~  \hat{s}^a ({\bf r}) {B}^a({\bf r})\;.
\end{eqnarray}
In Eq.\ \eqref{H_1} $\mu_B = 1/2c$ is the Bohr magneton, we employ Einstein's convention to sum over repeated indices ($a = x, y, z$), and
a multiplication by a $2\times2$ identity matrix is implied for the terms which are  diagonal in spin space, i.e., $-i \nabla + \frac{1}{c} {\bf A}({\bf r})$ and $v({\bf r})$. The electron-electron interaction  is given by
\begin{align}
  \cW = \frac{1}{2} \iind{3}{r_1}{3}{r_2} \frac{:\FPd(\br_1) \FP(\br_1)\FPd(\br_2) \FP(\br_2):}{|\br_1 - \br_2|} ~, \label{W}
\end{align}
where $:\cdots:$ denotes the normal ordering of the two-component Pauli field operators $\FPd = (\FPupd \; \FPdownd)$;
finally,
\begin{subequations} \label{current}
\begin{equation}
\hat{n}({\bf r}) =  \hat{ \Psi}^{\dagger}({\bf r})  \hat{ \Psi}({\bf r})
\end{equation}
is  the particle-density operator and
\begin{equation}
\hat{s}({\bf r}) = \hat{ \Psi}^{\dagger}({\bf r}) {\sigma}^a \hat{ \Psi}({\bf r})
\end{equation}
\end{subequations}
is the $a$-th component of the spin-density operator. 
Expanding the square and making use of partial integration, one obtains
\begin{eqnarray}\label{H_2}
\hat{H} &=&  \int d^3r~ \hat{\tau}(\br)  + \cW \nonumber \\
&+& \frac{1}{c} \int d^3r~ {\hat {\bf j}}(\br)  \cdot {\bf A}(\br) + \frac{\mu_B}{2c} \int d^3r~ {\hat { \bf J}^a}(\br) \cdot {{\bf A}^a }(\br)  \nonumber \\
&+& \int d^3r~ \hat{n}({\bf r})\tilde{v}({\bf r})
+ \mu_B \int d^3r~  \hat{s}^a({\bf r})  \tilde{B}^a(\bf r)
\end{eqnarray}
where
\begin{subequations}
\begin{equation}
 \hat{\tau}  = \frac{1}{2} \left( \nabla \FPd  \right) \cdot \left( \nabla \FP \right)
\end{equation}
is the kinetic-energy-density operator,
\begin{equation}
\hat{{\bf j}}  = 
\frac{1}{2i} \left[ \hat{ \Psi}^{\dagger}  \nabla \hat{ \Psi}  - \left( \nabla \hat{ \Psi}^{\dagger} \right) \hat{ \Psi}\right] 
\end{equation}
is the paramagnetic-current operator, and 
\begin{equation}
\hat{{\bf J}}^a = 
\frac{1}{2i} \left[ \hat{ \Psi}^{\dagger} \sigma^a \nabla \hat{ \Psi}  -
\left( \nabla \hat{ \Psi}^{\dagger} \right)  \sigma^a \hat{ \Psi}\right] 
\end{equation}
is the paramagnetic-spin-current operator.
\end{subequations}
We have also defined
\begin{equation}
\tilde{v} =   v + \frac{1}{2c^2 }  \left[ {\bf A} \cdot {\bf A} + \frac{\mu^2_B}{4} {\bf A}^a \cdot {\bf A}^a \right]\;,
\end{equation}
\begin{equation}
\tilde{B}^a =  B^a +  \frac{1}{2 c^2}  {\bf A} \cdot {\bf A}^a\;.
\end{equation}

Given the external fields ${\bf A}$, ${\bf A}^a$, $v$, and ${B}^a$, 
the ground-state energy is the expectation value of $\hat{H}$ in the corresponding ground state $| \Psi \rangle$.
The ground-state energy can be determined by means of a constrained-search minimization~\cite{Levy:82,Lieb:83}:
\begin{align}\label{E}
E &= \min_{(n,~ {s}^a,~ {\bf j},~ { \bf J}^a)} \Big\{ F[n, {s}^a, {\bf j}, { \bf J}^a ]  \nonumber \\
&+  \frac{1}{c} \int d^3r~ {{\bf j}}(\br)  \cdot {\bf A}(\br) +  \frac{\mu_B}{2c} \int d^3r~ {{ \bf J}^a}(\br) \cdot {{\bf A}^a }(\br)  \nonumber \\
&+  \int d^3r~ {n}({\bf r})\tilde{v}({\bf r})
+ \mu_B \int d^3r~  {s}^a({\bf r})  \tilde{B}^a(\bf r) \Big\}
\end{align}
with
\begin{align}\label{F}
F[n,  {s}^a, {\bf j}, { \bf J}^a ] = \min_{ | \Psi'\rangle \rightarrow (n,~ {s}^a,~ {\bf j},~ { \bf J}^a)} \langle \Psi | \hat{T} + \hat{W} | \Psi \rangle 
\end{align}
where the inner minimization is carried out over all the  many-body  wave functions yielding the
prescribed  set of densities and the outer minimization is carried out with respect to all $N$-representable densities. 
Eq.\ \eqref{F} defines a universal density functionals, which is the direct generalization of the universal functional in standard DFT. 
Assuming that the same set of densities is both interacting and non-interacting $v$-representable, one can further decompose 
$F$
\begin{align}\label{FKS}
F[n,  {s}^a, {\bf j}, { \bf J}^a ] &= T_{\rm KS}[n,  {s}^a, {\bf j},  {\bf J}^a ] + E_H[n] + E_{\rm xc}[n, {s}^a,  {\bf j}, { \bf J}^a ] 
\end{align}
in terms of the Kohn-Sham (KS) kinetic energy (see below)
$T_{\rm KS}[n,  {s}^a, {\bf j}, { \bf J}^a ]$, the Hartree energy $E_{\rm H}[n] = \frac{1}{2} \int d^3r \int d^3r' ~ \frac{{n}({\bf r}) {n}({\bf r}')}{|\br - \br'| }$ and a
remainder, $E_{\rm xc}[n, {s}^a,  {\bf j}, { \bf J}^a ] $, which is the xc-energy functional. 
In this way,
the  problem of determining the ground-state energies is reformulated into devising practical and sufficiently accurate approximations  for $E_{\rm xc}$.
In order to simplify the notation in the following we redefine the external potentials $\mu_B {\bf B} \to {\bf B}$ and $\frac{\mu_B}{2}  {\bf A}^a \to {\bf A}^a$.

The KS equations in SCDT have the form of single-particle Pauli equations
\begin{equation}\label{KSeq}
\left[ \frac{1}{2}\left( -i   \nabla + \frac{1}{c} { \mathbfcal{A} }_{\rm KS}  \right)^2 +  {\cal V}_{\rm KS}  \right] \Phi_\mu = \varepsilon_k \Phi_k
\end{equation}
where
\begin{equation}
{ \mathbfcal{A} }_{\rm KS} =\left( {\bf A} + {\bf A}_{{\rm xc}} \right) +  \sigma^a \left( {\bf A}^a + {\bf A}^a_{{\rm xc}} \right)\;,
\end{equation}
\begin{align}
{\cal V}_{\rm KS} &= \left( v + v_{\rm H}+ v_{\rm xc} \right) + \sigma^a\left( B^a + B^a_{\rm xc} \right) \nonumber \\
&+ \frac{1}{2c^2} \left[ \left(  {\bf A} + \sigma^a {\bf A}^a \right)^2-  {\mathbfcal A}^2_{\rm KS} \right]\;,
\end{align}
in which
$
{\frac{1}{c}}{\bf A}_{\rm xc} = \frac{\delta E_{\rm xc}}{\delta {\bf j}(\br)}$
is the Abelian xc-vector potential,
$
{\frac{1}{c}}{\bf A}^a_{\rm xc} = \frac{\delta E_{\rm xc}}{\delta { \bf J}^a(\br)}
$
is the $a$-th component of the non-Abelian xc-vector potential,
$
{B}^a_{\rm xc} = \frac{\delta E_{\rm xc}}{\delta {s^a}(\br)}
$
is the $a$-th component of the xc-magnetic potential (Zeeman field),
$
 v_{\rm xc} = \frac{\delta E_{\rm xc}}{\delta n(\br)}
$
is the xc-scalar potential, and $v_H(\br) = \int d\br \frac{n(\br')}{|\br - \br'|}$ is the usual Hartree potential.

A fundamental properties of $E_{\rm xc}$ is its  invariance under general U(1)$\times$SU(2) gauge transformations.
We recall that a local U(1) transformation, $U({\bf r}) $, is defined by
\begin{eqnarray}\hat{\Psi} ({\bf r}) \rightarrow  \hat{\Psi}' ({\bf r})   =   \exp\left[ \frac{i}{c}  \chi(\bf r)  \right] \hat{\Psi} ({\bf r}) \end{eqnarray}
where $ \chi(\bf r) $ is a scalar function of the position, 
and a local SU(2) transformation is defined by
\begin{eqnarray} \hat{\Psi} ({\bf r}) \rightarrow  \hat{\Psi}' ({\bf r})  &=& \exp\left[ \frac{i}{c} \lambda^a({\bf r}) { \sigma}^a \right]  \hat{\Psi} ({\bf r}) \nonumber \\ 
&=& U_{\rm S}({\bf r})  \hat{\Psi} ({\bf r}) \end{eqnarray}
where $\lambda^a(\bf r)$ are the components of a  vector function of the position.
A detailed analysis of gauge  transformations in SCDFT framework is presented in Ref.s [\onlinecite{VignaleRasolt:88,Bencheikh:03}]. 
Note that neither $F[n, {s}^a,  {\bf j}, { \bf J}^a ]$ nor $T_{\rm KS}[n, {s}^a, {\bf j}, { \bf J}^a ]$ are invariant, but 
they transform in the same way because the KS system has the same densities $n$, ${s}^a$,  ${\bf j}$, and ${ \bf J}^a$ as the interacting system.
As a result 
\begin{equation}\label{tExc}
E_{\rm xc}[n', {s'}^a,  {\bf j}', {{\bf j}'}^a ]= E_{\rm xc}[n, {s}^a,  {\bf j}, { \bf J}^a ]\;,
\end{equation}
where
\begin{subequations}\label{tDens}
\begin{equation}
n  \rightarrow  n' = n
\end{equation}
\begin{equation}
{s}^a \rightarrow {s'}^a = R^{ab} {s}^b
\end{equation}
\begin{align}
{\bf j} \rightarrow {\bf j}' = {\bf j} + \frac{1}{c} n \nabla \chi - \frac{i}{2} s^a  {\rm Tr} \left( \sigma^a U^{\dagger}_{\rm S} \nabla  U_{\rm S}  \right)
\end{align}
and
\begin{align}
{ \bf J}^a \rightarrow {{\bf J}'}^a  = R^{ab} \left[ {\bf J}^b + \frac{1}{c}  {s}^b   \nabla \chi
 - \frac{i}{2} n  {\rm Tr} \left( \sigma^b U^{\dagger}_{\rm S} \nabla  U_{\rm S} \right) \right]\;.
\end{align}
\end{subequations}
where the matrix $R^{ab}$ is a $3 \times 3$ matrix describing a rotation in $\mathbb{R}^3$ around $\hat{\lambda}$ -- the unit vector in the direction of $\lambda^a$ -- by an angle $\varphi=-2 |\lambda| / c$ and {\rm Tr} is the trace taken with respect to spin indices.

The transformation of the xc-fields  can be readily  deduced by combining the invariance of $E_{\rm xc}$ -- as expressed  in Eq.~(\ref{tExc}) -- 
with the transformations of the densities, as given in Eq.~(\ref{tDens}). For completeness, 
we have
\begin{subequations}\label{tFields-KS}
\begin{equation}\label{tAxc}
{\bf A}'_{\rm xc} = {\bf A}_{\rm xc}
\end{equation}
\begin{equation}\label{tNAAxc}
{{\bf A}'}^a_{\rm xc} =  R^{ab} {{\bf A}}^b_{\rm xc} 
\end{equation}
\begin{equation}\label{tBxc}
{{B}'}^a_{\rm xc} =  R^{ab} 
\left[ B^b_{\rm xc}  + \frac{i}{2} {\bf A}_{{\rm xc}} \cdot {\rm Tr}  \left( \sigma_b U^{\dagger}_{\rm S} \nabla U_{\rm S} \right)  - {\bf A}^b_{\rm xc} \cdot \nabla \chi \right]
\end{equation}
\begin{equation}\label{tvxc}
v'_{\rm xc} = v_{\rm xc} - \frac{1}{c} {\bf A}_{\rm xc} \cdot \nabla \chi + \frac{i}{2} {\bf A}^a_{\rm xc} \cdot {\rm Tr} \left( \sigma^a U^{\dagger}_{\rm S} \nabla U_{\rm S}  \right)\;.
\end{equation}
\end{subequations}
It is apparent that while ${\bf A}_{{\rm xc},{\mu }}$ is invariant, $B^a_{\rm xc}$, ${{\bf A}'}^a_{\rm xc}$, and $v_{\rm xc}$  are not invariant, but {\it covariant}~\cite{note2}.  Note that
even in the case of a restricted U(1) transformation,
$B^a_{\rm xc}$, ${\bf A}_{{\rm xc}, \mu}$, and $v_{\rm xc}$ {\em do not} behave as standard Maxwellian fields and, in general,
${B}^a_{\rm xc} \ne \left[ \nabla \times {\bf A}_{{\rm xc}} \right]_a$.
The xc-fields in SCDFT should be regarded as  some  effective Yang-Mills fields.

Although the xc-fields are not generated by any physical field equations, 
they are bound to satisfy compatibility relations -- having the form of conservation laws~\cite{Bencheikh:03} -- due to the invariance of $E_{\rm xc}$, i.e.,
\begin{subequations}
\begin{equation}\label{c1_2}
\partial_\mu   \left[ n  {A}_{{\rm xc},\mu}   + \vec{s} \cdot   \vec{A}_{{\rm xc},\mu} \right] = 0
\end{equation}
and 
\begin{align}\label{c2_2}
\frac{1}{2c} \partial_\mu \cdot \left[ {A}_{{\rm xc},\mu} \vec{s} + n \vec{A}_{{\rm xc},\mu} \right]   =  \frac{1}{c}  \vec{A}_{{\rm xc},\mu} \times \vec{ j}_\mu  + \vec{B}_{\rm xc} \times \vec{s}\;.
\end{align}
\end{subequations}
In writing these expressions, we have denoted vectors in spin space (i.e., vectors with a non-Abelian index) with an arrow to highlight the torques due to  $\vec{A}_{{\rm xc},\mu}$ and  $\vec{B}_{\rm xc}$ on the 
right hand side of Eq.~(\ref{c2_2}). 
The KS system of SCDFT reproduces the interacting paramagnetic currents but may not reproduce the diamagnetic currents. Whatever the difference between the KS and interacting diamagnetic currents is,
Eq.~(\ref{c1_2}) and Eq.~(\ref{c2_2}) ensure that 
the stationarity conditions for the particle and spin densities are not violated: $\vec{A}_{{\rm xc},\mu}$ and $ \vec{B}_{\rm xc} $ can balance any non-vanishing xc-divergence-like contribution.

There is one non-trivial case in which we can see that the solution of the KS equations in SCDFT reduces to the solution of the analogous equations in SDFT.
First, note that, to have ${A}_{{\rm xc},\mu} = 0$ and $\vec{A}_{\rm xc,\mu} = 0$,
the torque of $\vec{B}_{\rm xc}$ must vanish as well [see Eq.~(\ref{c2_2})].  Yet $\vec{B}_{\rm xc}$ can be non-vanishing, if it
is  parallel to the spin density at every point in space.
If the external non-Abelian vector potential is also vanishing, we are then in a situation in which SDFT applies rigorously. 
Thus, in this case, we can conclude that  $ {v}_{\rm xc} = {v}^{\rm SDFT}_{\rm xc} $ and  $ \vec{B}_{\rm xc} = \vec{B}^{\rm SDFT}_{\rm xc} $.

\section{Short-range behavior of exchange-only pair-correlation functions}\label{Sec3}

Importing the standard Local-Spin-Density Approximation (LSDA)  in SCDFT does not allow us to fully exploit the power of SCDFT, as
the LSDA only depends on the magnitude of the spin density and the particle density.
Moreover, the LSDA is insensitive to strong  inhomogeneities and long-range interactions.
Exact exchange would be an obvious, more sophisticated choice~\cite{RG06}, but
its combination with suitable correlation functionals may require more involved computational approaches.
Generalized-Gradient Approximations (GGAs) and, more recently, Meta-GGAs~\cite{MGGAS} (MGGA)
-- either stand-alone or combined with the Hartree-Fock method into hybrids -- are the gold standard in modern DFT calculations.
Here, we report an analysis that points to the fact that MGGAs are ideal forms to satisfy U(1)$\times$SU(2) gauge-invariance while fulfilling
other exact properties of the underlying pair-correlation functions.

We begin by reviewing known definitions about the so-called exchange hole.  Assuming, as it is normally done, that the KS states are in the form of single Slater determinants, the exchange energy can be expressed as
\begin{equation}\label{Ex_1}
E_{\rm x} = - \frac{1}{2} \int~d^3r \int d^3r' \frac{{\rm Tr} \left\{ \Gamma({\bf r},{\bf r}')\Gamma({\bf r}',{\bf r}) \right\}}{ | {\bf r} - {\bf r}' | }
\end{equation}
where
\begin{equation}\label{gamma}
\Gamma({\bf r},{\bf r}') = \sum_{k=1}^{N} \Phi_k({\bf r})\Phi^\dagger_k({\bf r}')
\end{equation}
is  the   one-body-reduced-spin density matrix obtained from the occupied spinors, which are solutions of Eq.~(\ref{KSeq}).
$\Gamma({\bf r},{\bf r}')$ is a 2$\times$2 matrix in spin space.
$E_{\rm x}$ is evidently invariant under  general U(1) $\times$ SU(2) gauge transformations.

$E_{\rm x}$ can be usefully expressed in terms of the x-hole function, for which a convenient definition, applicable to non-collinear spin states is
\begin{equation}\label{hx}
h_{\rm x}({\bf r},{\bf r}') := - \frac{ {\rm Tr} \left\{ \Gamma({\bf r},{\bf r}')\Gamma({\bf r}',{\bf r}) \right\} }{n({\bf r}) }\,.
\end{equation}
In practice, the spherical average 
\begin{equation}\label{shx}
h_{\rm x}({\bf r},u) := \frac{1}{4\pi} \int d\Omega_{\bf u}~ h_{\rm x}({\bf r},{\bf r} + {\bf u})
\end{equation}
is what really matters to the end of the calculation of the exchange energies.
 
In Eq.~(\ref{shx}), the integration is carried out with respect to the solid angle $\Omega_{\bf u}$ formed by ${\bf r}$ and ${\bf u}$;
 ${\bf r}$ is the so-called reference position.
Thus, we rewrite
\begin{equation}\label{Ex_2}
E_{\rm x} = 2 \pi  \int d^3r~ n({\bf r}) \int du\;\; u\; h_{\rm x}({\bf r},u)\;.
\end{equation}

Taylor-expanding $ h_{\rm x}  ({\bf r}, u ) $  for small inter-particle separations $u$, we find
\begin{equation}\label{TE-Hx}
 h_{\rm x}    ({\bf r}, u ) = - \frac{n({\bf r})}{2} \left( 1 + \frac{ s^a({\bf r})  s^a({\bf r})}{n^2({\bf r})} \right) \\
-  C^{\rm nc}_{\rm x}({\bf r}) u^2 + \cdot \cdot \cdot
\end{equation}
where
\begin{widetext}
\begin{align}\label{C-Hx}
C^{\rm nc}_{h_{\rm x}}  &= \frac{1}{3}
  \left\{ \left[   \left(\tau - \frac{{\bf j} \cdot {\bf j} }{ 2n } \right)   - \left(   \frac{\nabla^2 n}{4}  + \frac{\nabla n \cdot \nabla n  }{8 n } \right) \right] 
+ \left[  \left(  \frac{ {s}^{a}   {\tau}^{a}  }{n}
- \frac{ {\bf  J}^{a}  \cdot {\bf  J}^{a}  }{ 2 n } \right) - \left( \frac{ {s}^{a}  \cdot  \nabla^2 {s}^{a}    }{ 4 n } +  \frac{ ( \nabla {s}^{a}  ) \cdot  ( \nabla {s}^{a}  )}{8 n } \right)  \right] \right\}
\end{align}
\end{widetext}
is the curvature of the x-hole: the superscript ``nc'' emphasizes that this expression differs
from the analogous quantity derived earlier for  spin-unpolarized or spin-polarized {\em but}  globally collinear states. 
By admitting complex-valued single-particle spinors,  which is the natural state of affairs in non-collinear spin systems, we find terms that depend  not only on the usual  kinetic energy density, 
\begin{equation}
\tau({\bf r}) = \frac{1}{2} \sum_{k=1}^{N} \Big( \partial_{\mu} \Phi_k^{\dagger}({\bf r}) \Big) \Big( \partial_{\mu} \Phi_k({\bf r}) \Big)\;,
\end{equation}
but also on the  {\it spin-kinetic energy density} defined as
\begin{equation}\label{SpinKineticEnergy}
{\tau}^{a} ({\bf r}) = \frac{1}{2} \sum_{k=1}^N \Big(  \partial_\mu \Phi^{\dagger}_k({\bf r}) \Big) {\sigma}^{a} \Big( \partial_\mu \Phi_k({\bf r}) \Big)\;.
\end{equation}
Eq.~(\ref{TE-Hx}) and Eq.~(\ref{C-Hx}) provide the exact short-range behavior of the  x-hole function in presence of particle- and spin-currents for  
non-relativistic non-collinear states.
The  gauge-invariance of the expansion coefficients in Eq.~(\ref{TE-Hx}) is obvious:  a scalar function has been expanded with respect to a scalar variable.
But one may also verify this property by direct inspection. 

The on-top x-hole (i.e., the first term on the right hand side of  Eq.~(\ref{TE-Hx})) already provides us with an explicit indication on how the on-top x-hole of an 
existing DFT functional should be modified to admit an extra {\em non-empirical} dependence on U(1) $\times$ SU(2) gauge-invariant  quantities; simply, 
$n \to n \left( 1 + \frac{ {s}^a {s}^a }{ n^2 } \right)$.

At the level of the  curvature of the x-hole [Eq.~(\ref{C-Hx})], notice that 
the  combination $\left[   \left(\tau - \frac{{\bf j} \cdot {\bf j} }{ 2n } \right)   - \left(   \frac{\nabla^2 n}{4}  + \frac{\nabla n \cdot \nabla n  }{8 n } \right) \right]$ 
is already known to be a U(1) gauge-invariant quantity: the difference, here, is that  all the quantities are evaluated on fully {\em non-collinear} two component spinors.
Thus, this expression is {\em not}  SU(2) gauge invariant. The contribution
$\left[  \left(  \frac{ {s}^{a}   {\tau}^{a}  }{n}
- \frac{ {\bf  J}^{a}  \cdot {\bf  J}^{a}  }{ 2 n } \right) - \left( \frac{ {s}^{a}  \cdot  \nabla^2 {s}^{a}    }{ 4 n } +  \frac{ ( \nabla {s}^{a}  ) \cdot  ( \nabla {s}^{a}  )}{8 n } \right)  \right] $
is  essential to get the full invariance. A detailed discussion of the transformation of each term is presented in Appendix~\ref{Appendix}. 

Therefore, the minimal substitution
\begin{align}\label{eta1}
\tau \rightarrow    \tilde{\tau} &= \left(  \tau - \frac{{\bf j} \cdot {\bf j} }{ 2n }  \right) + \left(  \frac{ {s}^{a}   {\tau}^{a}  }{n}
- \frac{ {\bf  J}^{a}  \cdot {\bf  J}^{a}  }{ 2 n } \right)  \nonumber \\
&- \left( \frac{ {s}^{a}  \cdot  \nabla^2 {s}^{a}    }{ 4 n } +  \frac{ ( \nabla {s}^{a}  ) \cdot  ( \nabla {s}^{a}  )}{8 n } \right)\;
\end{align}
can be used to transform a DFT-MGGA form into a SCDFT-MGGA form. 

In practical applications it is often desirable to eliminate the laplacian terms, which may be difficult to evaluate numerically.   To accomplish this, one substitutes the x-hole, expressed in terms of ${C}^{\rm nc}_{h_{\rm x}}$, 
into the expression for the exchange
energy density and performs an integration by parts to show that ${C}^{\rm nc}_{h_{\rm x}}$ is actually equivalent (as far as the calculation of the exchange energy is concerned) to
 \begin{widetext}
\begin{align}\label{tildeC-Hx}
\bar{C}^{\rm nc}_{h_{\rm x}} =  \frac{1}{3}
\left[ \left(  \tau - \frac{{\bf j} \cdot {\bf j} }{ 2 n }   + \frac{\nabla n \cdot \nabla n  }{8 n  }  \right)
+ \left( \frac{{s}^{a}   {\tau}^{a}}{n}   - \frac{ {\bf  J}^{a}  \cdot {\bf  J}^{a}  }{ 2n  }  +   \frac{ ( \nabla {s}^{a}  ) \cdot  ( \nabla {s}^{a}  )}{8 n }   \right) \right]\;,
\end{align}
\end{widetext}
which no longer contains the laplacian operator.  Obviously,  the form of $\bar{C}^{\rm nc}_{h_{\rm x}}$ implies the minimal substitution
\begin{align}\label{eta2}
 \tau \rightarrow   \tilde{\tau}  = \left(  \tau - \frac{{\bf j} \cdot {\bf j} }{ 2n }  \right)  + \left( \frac{{s}^{a}   {\tau}^{a}}{n}   - \frac{ {\bf  J}^{a}  \cdot {\bf  J}^{a}  }{ 2n  }  \right)  +   \frac{ ( \nabla {s}^{a}  ) \cdot  ( \nabla {s}^{a}  )}{8 n }\;.
\end{align} 

Alternatively,  functional approximations are also constructed  working directly at the level of the one-body  density matrix
\begin{equation}
Q_{\rm x}({\bf r},{\bf r}') = {\rm Tr} \left\{ \Gamma({\bf r},{\bf r}')\Gamma({\bf r}',{\bf r}) \right\}\;,
\end{equation}
in terms of which the  exchange energy can be expressed as
\begin{eqnarray} \label{Ex-Qx}
E_{\rm x} &=& - \frac{1}{2} \int d^3 r \int d^3 u~\frac{ Q_{\rm x}({\bf r} + {\bf u}/2,{\bf r} - {\bf u}/2) }{ u} \nonumber \\
&=&   - 2\pi \int d^3r \int  u d u~    Q_{\rm x}  ({\bf r}, u ) 
\end{eqnarray}
where the vector positions are expressed with respect to the coordinates of the center of mass.
Taylor-expanding $ Q^{\rm nc}_{\rm x}  ({\bf r}, u )$ with respect to $u$, we obtain
\begin{equation}\label{TE-Qx} 
 Q_{\rm x}  ({\bf r}, u )  =   \frac{ \left[ n^2({\bf r}) +{s}^a({\bf r}) {s}^a({\bf r}) \right]}{2} 
+ C^{\rm nc}_{Q_{\rm x}}({\bf r})  u^2 + \cdot \cdot \cdot
\end{equation}
where
\begin{widetext}
\begin{eqnarray}\label{C-Qx}
C^{\rm nc}_{Q_{\rm x}} =   \frac{1}{3} 
\left[ \left(   n \tau - \frac{ {\bf j} \cdot {\bf j} }{2}    - \frac{ n \nabla^2 n}{8}  \right) 
+  \left( {s}^a {\tau}^a- \frac{ {\bf  J}^{a} \cdot { \bf J}^a }{2} - \frac{ {s}^a   \nabla^2 {s}^a   }{ 8 }  \right) \right] \;.
\end{eqnarray}
\end{widetext}
$C^{\rm nc}_{Q_{\rm x}}$ generalizes the known DFT expression to SCDFT.

Therefore,  in extending a DFT-MGGA form based on 
 the short-range of  $ Q_{\rm x}  $, we can proceed by performing  two minimal substitutions: the first one,  $n^2 \to \left( n^2  +  {s}^a {s}^a  \right) $, has to be carried out only
 at the level of the on-top quantities; the second one is performed at the level of $\tau$
\begin{eqnarray}\label{eta3}
\tau   \to    \tilde{\tau}  = \left(  \tau - \frac{{\bf j} \cdot {\bf j} }{ 2n }  \right)  +  \left( {s}^a {\tau}^a- \frac{ {\bf  J}^{a}  \cdot { \bf J}^a }{2}\right) - \frac{ {s}^a   \nabla^2 {s}^a   }{ 8 }.\nonumber\\
\end{eqnarray}

Again, an intermediate integration by parts for the calculation of the  exchange energies 
yields an alternative form of the curvature factor:
\begin{equation}\label{tildeQ-Hx} 
C^{\rm nc}_{Q_{\rm x}} \rightarrow \bar{C}^{\rm nc}_{Q_{\rm x}} = n \bar{C}^{\rm nc}_{h_{\rm x}}\; 
\end{equation}
which, modulo an overall multiplication by the particle density,  implies the same minimal substitution as in Eq.~(\ref{eta2}).\\

\section{Construction of functionals}\label{Sec4}
We conclude our analysis by constructing two new exchange-energy functionals based on
existing forms and proposing them for immediate use.

First, let us consider the construction of the BR89\cite{BR89}: this was derived allowing only globally collinear spin polarization.
In this approach, the x-hole of the Hydrogen atom is turned into a general model
by introducing
two position-dependent  ``parameters" $p_1({\br})$ and $p_2({\br})$
\begin{widetext}
\begin{equation}
h^{{\rm model}}_{ {\rm x}}({\bf r},u) = - \frac{ p_1({\bf r}) }{ 16 \pi p_2({\bf r}) u } 
\left[  p_1({\bf r}) \left( |p_2({\bf r}) - u| + 1  \right)e^{-p_1({\bf r})|p_2({\bf r})-u|} 
- p_1({\bf r}) \left( |p_2({\bf r}) + u| + 1  \right)e^{-p_1({\bf r})|p_2({\bf r})+u|} \right]
\end{equation}
\end{widetext}
to be chosen in such a way  to reproduce the short-range behavior of the x-hole of an $N$-electron system.

Eq.~(\ref{TE-Hx}) and Eq.~(\ref{C-Hx}) allow us to readily generalize this 
procedure to the non-collinear current-carrying states. As a result, the
$p_1({\br})$ and $p_2({\br})$ must be determined by solving the equations
\begin{subequations}
\begin{equation}
p_1^3 e^{-p_1p_2}  = 4 \pi n \left( 1 + \frac{s^a s^a}{n^2} \right)
\end{equation}
\begin{equation}
p_1^2 p_2 - 2 p_1=  12 b \frac{C^{\rm nc}_{h_{\rm x}}}{n\left( 1 + \frac{s^a s^a }{n^2} \right)}\;.
\end{equation}
\end{subequations}

As a second example, let us consider the approximations based on a Gaussian re-summation 
of the  short-range behavior of  $Q_{\rm x}  ({\bf r}, u )$.
For closed-shell systems (i.e., vanishing spin polarization), Lee and Parr~\cite{LP87} find
\begin{equation}
E^{\rm G}_{\rm x}  = - \pi \int d^3r~ n^2({\bf r}) \beta({\bf r})\;,
\end{equation}
For the sake of simplicity, here, we are not considering a more sophisticated form which would satisfy particle-number normalization for any system.
In view of Eq.~(\ref{Ex-Qx}) and Eq.~(\ref{TE-Qx}), the extension to
non-collinear spin-polarized current-caring states is readily obtained upon the 
substitutions
\begin{subequations}
\begin{equation}\label{LP1}
\beta \rightarrow  -\frac{1}{2} \frac{n^2 + {s}^a {s}^a }{{C}^{\rm nc}_{Q_{\rm x}}}\;
\end{equation}
and
\begin{equation}\label{LP2}
n^2 \rightarrow (n^2 + {s}^a {s}^a )\;.
\end{equation}
\end{subequations}
Finally, we note that ${C}^{\rm nc}_{Q_{\rm x}}$ is 
positive for single-particle states, for states with vanishing spin-currents, for the spin-spirals of the uniform gas,
but otherwise the question remains open.

\section{Conclusions}\label{Sec5}
This work opens the way to the extension of time-proven semi-local exchange-correlation energy functionals as well as to the derivation of novel 
approximations designed  to deal with non-collinear spin structures which are subject of great and large ongoing  interest. Specifically,
we have introduced non-empirical U(1)$\times$SU(2) gauge-invariant  building blocks  
which are, in principle, ideally suited for dealing with (static) spin-fluctuation  of {\em strongly} inhomogeneous states.
Our results  show how the gradients of the spin-density should be combined with the spin-kinetic-energy density -- an
information which should be  relevant even in devising semi-local forms within standard Spin-DFT -- and
the Kohn-Sham paramagnetic (spin-)currents appear as {\em explicit} ingredients as well.
Thus, we have  provided examples of  extension of existing exchange-only functional forms. 
On passing, we have also illustrated some of the exact fundamental features of the exact exchange-correlation fields.

\begin{acknowledgments}
S.P.  was  supported by the European Community through the FP7's Marie-Curie Incoming-International Fellowship, Grant agreement No. 623413.
G. V. acknowledges support from DOE Grant DE-FG02-05ER46203. 
F. G. E. has received funding from the European Union's Framework Programme for Research and Innovation Horizon 2020 (2014--2020) under the Marie Sk{\l}odowska-Curie Grant Agreement No. 701796.
\end{acknowledgments}

\begin{appendix}

\section{Derivation of the transformation laws} \label{APP:U1SU2TransformationLaws}\label{Appendix}

The expansions worked out in the main text have the advantage to spare us from the burden to explicitly deal with  the  transformations
of the densities under general gauge transformations -- this is because we could consistently deal only with scalars.
Nevertheless, in this appendix, we report
the derivation of the transformation laws of the considered densities and their salient combinations.
This should offer thorough clarifications and further insights.

We will use the following notational convention: Spatial indices are denoted by subscripts, spin indices by superscripts, and repeated indices are summed. 
Furthermore, the dependence on the position $\br$ is implied. We are interested in obtaining the transformation laws for the following
densities: 
\begin{subequations}
  \begin{align}
    n & = \POd \PO ~, \label{n} \\
    s^a & = \POd \sigma^a \PO ~, \label{s} \\
    j_\mu & = \tfrac{1}{2i} \big[\POd (\partial_\mu \PO ) - (\partial_\mu \POd) \PO \big] ~, \label{j} \\
    J_\mu^a & = \tfrac{1}{2i} \big[\POd \sigma^a (\partial_\mu \PO )
    - (\partial_\mu \POd) \sigma^a \PO \big] ~, \label{J} \\
    \tau & = \tfrac{1}{2} (\partial_\mu \POd) (\partial_\mu \PO ) ~, \label{t} \\
    \tau^a & = \tfrac{1}{2} (\partial_\mu \POd) \sigma^a (\partial_\mu \PO ) ~. \label{T}
  \end{align}
\end{subequations}
The combined U(1)$\times$SU(2) transformation is given by
\begin{align}
  U = \exp\Big[\tfrac{i}{c} \big(\chi + \lambda^a \sigma^a \big)\Big]
  = \exp\big[\tfrac{i}{c} \chi \big]
  \exp\big[\tfrac{i}{c} \lambda^a \sigma^a \big] ~, \label{U1SU2}
\end{align}
where we use that the U(1) and SU(2) transformation commute. This means that we can investigate the U(1) and SU(2) transformation laws separately. \\

{ \bf \emph{ U(1) transformation laws} }-- The density, $n$, and the spin magnetization, $s^a$, are trivially invariant under local U(1) transformations. 
It is straightforward to obtain the transformation laws for the remaining 
densities, i.e.,

\begin{subequations} \label{U1Transformation}
  \begin{align}
    n & \to n \\
    s^a & \to s^a  \\
    j_\mu & \to j_\mu + \tfrac{1}{c} (\partial_\mu \chi) n ~, \label{jU1} \\
    J_\mu^a & \to J_\mu^a + \tfrac{1}{c} (\partial_\mu \chi) s^a ~, \label{JU1} \\
    \tau & \to \tau + \tfrac{1}{c} (\partial_\mu \chi) j_\mu
    + \tfrac{1}{2c^2} (\partial_\mu \chi) (\partial_\mu \chi) n ~, \label{tU1} \\
    \tau^a & \to \tau^a + \tfrac{1}{c} (\partial_\mu \chi) J_\mu^a
    + \tfrac{1}{2c^2} (\partial_\mu \chi) (\partial_\mu \chi) s^a ~. \label{TU1}
  \end{align}
\end{subequations}
It follows directly that the combinations 
\begin{subequations}
  \begin{align}
    & n t - \tfrac{1}{2} j_\mu j_\mu ~, \label{U1invariant1} \\
    & s^a \tau^a - \tfrac{1}{2} J_\mu^a J_\mu^a ~, \label{U1invariant2}
  \end{align}
\end{subequations}
are invariant under local U(1) transformations.\\

{ \bf \emph{ Infinitesimal SU(2) transformation laws} }-- 
The density is trivially invariant under local SU(2) transformations. The spin magnetization, however, is not invariant. Using that
\begin{align}
  U_S = \exp\big[\tfrac{i}{c} \lambda^a \sigma^a \big]
  = \cos[\lambda/c] + i \sin[\lambda/c] \hat{\lambda}^a \sigma^a ~, \label{US}
\end{align}
where $\lambda$ is the magnitude and $\hat{\lambda}^a$ is the unit vector in the direction of the vector $\lambda^a$.
We recall 
\begin{align}
  U_S^\dagger \sigma^a U_S & = \cos[2 \lambda / c] \sigma^a 
  - \sin[2 \lambda / c] \epsilon^{abc} \hat{\lambda}^b \sigma^c \label{US_R}~\nonumber \\
  & \phantom{=} {} + \big(1 - \cos[2 \lambda / c] \big) \hat{\lambda}^a \hat{\lambda}^b \sigma^b
  = R^{ab} \sigma^b ~.  
\end{align}\label{EqR}
The matrix $R^{ab}$ is a $3 \times 3$ matrix describing a rotation in $\mathbb{R}^3$ around the direction $\hat{\lambda}$ by an angle $\varphi=-2\lambda / c$. It follows that the spin magnetization transforms as
\begin{align}
  s^a \to R^{ab} s^b ~. \label{sSU2}
\end{align}

Before embarking on the derivation of the transformation laws for the other 
densities, we consider the case of \emph{infinitesimal} transformations. This means that we can approximate
\begin{align}
  U_S \approx 1 + \tfrac{i}{c} \lambda^a \sigma^a ~. \label{US_infinitesimal}
\end{align}
Keeping terms up to the first order in $\lambda$, we arrive at
\begin{subequations} \label{SU2TransformationInf}
  \begin{align}
    n & \to n \\
    s^a & \to s^a - \tfrac{2 \lambda}{c} \epsilon^{abc} \hat{\lambda}^b s^c ~, \label{sSU2inf} \\
    j_\mu & \to j_\mu + \tfrac{1}{c} (\partial_\mu \lambda^a) s^a ~, \label{jSU2inf} \\
    J_\mu^a & \to J_\mu^a - \tfrac{2 \lambda}{c} \epsilon^{abc} \hat{\lambda}^b J_\mu^c
    + \tfrac{1}{c} (\partial_\mu \lambda^a) n ~, \label{JSU2inf} \\
    \tau & \to \tau + \tfrac{1}{c} (\partial_\mu \lambda^a) J_\mu^a ~, \label{tSU2inf} \\
    \tau^a & \to \tau^a - \tfrac{2 \lambda}{c} \epsilon^{abc} \hat{\lambda}^b \tau^c
    + \tfrac{1}{c} (\partial_\mu \lambda^a) j_\mu \nonumber \\
    & \phantom{\to} {}
    - \tfrac{1}{2 c} \epsilon^{abc} (\partial_\mu \lambda^b) (\partial_\mu s^c) ~. \label{TSU2inf}
  \end{align}
\end{subequations}

 It is straightforward  to verify that neither the quantity in Eq.~(\ref{U1invariant1}) nor the combination in Eq.~(\ref{U1invariant2})
are invariant under these transformations. Yet, the overall invariance  of the x-only curvatures given in the main text can be now explicitly
verified for arbitrary infinitesimal U(1) and SU(2) transformations in a direct calculations. 
This task may be further simplified by considering the identities:
$\nabla s^a  \cdot \nabla s^a =   \frac{ \nabla^2 }{2}  s^a  s^a  -  s^a   \nabla^2 s^a $ and
$\nabla n  \cdot \nabla n =   \frac{ \nabla^2 }{2}  n^2  -  n   \nabla^2 n $.

Furthermore, the behavior of the basic densities under the same transformations 
suffices to establish the compatibility conditions \eqref{c1_2} and \eqref{c2_2} presented in the main text.
This is achieved by using the fact that  the xc energy is
invariant under the corresponding U(1) $\times$ SU(2) transformations.

{ \bf \emph{ Finite SU(2) transformation laws} }--
In principle, it is sufficient to establish invariance under infinitesimal SU(2) transformation, as an arbitrary finite SU(2) transformation can be represented as a sequence of infinitesimal transformations. However, the derivation of the transformation laws of the xc potentials requires  knowledge  of the  explicit transformations of the basic densities under 
arbitrary {\em finite} SU(2) transformations. Moreover, we here intend to spell out  the finite transformation of the spin-kinetic-energy density as well.

The difficulty in obtaining finite transformation laws is due to the fact that the generators of the SU(2) group, i.e., the Pauli matrices $\sigma^a$, do not commute. Accordingly, we have to keep in mind that
\begin{align}
  \partial_\mu U_S \neq \tfrac{i}{c} (\partial_\mu \lambda^a) \sigma^a U_S
  \neq \tfrac{i}{c} U_S (\partial_\mu \lambda^a) \sigma^a ~. \label{dUSwrong}
\end{align}
Instead, using the very definition of the directional derivative, we have
\begin{align}
  \partial_\mu U_S & \equiv \lim_{\epsilon \to 0} \tfrac{1}{\epsilon} 
  \Big( \exp\big[\tfrac{i}{c} \lambda^a(\br + \epsilon \hat{k}) \sigma^a \big]
  - \exp\big[\tfrac{i}{c} \lambda^a(\br) \sigma^a \big] \Big) \nonumber \\
  & = \lim_{\epsilon \to 0} \tfrac{1}{\epsilon} 
  \Big( \exp\big[\tfrac{i}{c} (\lambda^a + \epsilon \partial_\mu \lambda^a) \sigma^a \big]
  - \exp\big[\tfrac{i}{c} \lambda^a(\br) \sigma^a \big] \Big) \nonumber \\
  & = \left. \partial_\epsilon
  \exp\big[\tfrac{i}{c} (\lambda^a + \epsilon \partial_\mu \lambda^a) \sigma^a \big]
  \right|_{\epsilon = 0} ~. \label{dUS}
\end{align}
Now, we can use the identity
\begin{align}
  \partial_\gamma e^{\gamma(\hat{A} + \epsilon \hat{B})}
  & = e^{\gamma(\hat{A} + \epsilon \hat{B})} \indll{\gamma'}{0}{\gamma}
  e^{-\gamma'(\hat{A} + \epsilon \hat{B})} \hat{B} e^{\gamma'(\hat{A} + \epsilon \hat{B})}
  \nonumber \\ 
  & = \indll{\gamma'}{0}{\gamma} e^{\gamma'(\hat{A} + \epsilon \hat{B})} \hat{B}
  e^{-\gamma'(\hat{A} + \epsilon \hat{B})} e^{\gamma(\hat{A} + \epsilon \hat{B})} 
  ~, \label{MerminIdentity}
\end{align}
-- which is readily verified by noting that both sides fulfill the same first-order differential equation in $\gamma$ and vanish for $\gamma=0$ -- and write
\begin{align}
  U_S^\dagger \big(\partial_\mu U_S \big) & = 
  \tfrac{i}{c} (\partial_\mu \lambda^a) \indll{\gamma}{0}{1}
  \exp\!\!\big[-\tfrac{i \gamma}{c} \lambda^b \sigma^b \big] \sigma^a
  \exp\!\!\big[\tfrac{i \gamma}{c} \lambda^c \sigma^c \big] \nonumber \\
  & = \tfrac{i}{c} (\partial_\mu \lambda^a) \Lambda^{ab} \sigma^b ~, \label{UdU}
\end{align}
where we have introduced
\begin{align}
  \Lambda^{ab} = \indll{\gamma}{0}{1} R^{ab}(\gamma) ~. \label{Lambda}
\end{align}
The $3 \times 3$ matrix $\Lambda^{ab}$ is the uniform average over the rotation matrices around axis $\hat{\lambda}^a$ with rotation angles $\varphi \in [0, -2\lambda / c]$. Note that $\Lambda^{ab}$ \emph{is not} a rotation matrix itself. Using the group properties of the rotation matrices, one can easily verify the relation 
\begin{align}
  R^{ac} \Lambda^{bc} = \Lambda^{ab} ~, \label{RL_L}
\end{align}
which is an extremely useful identity as it shows how the rotation matrix $R^{ab}$ acts on the matrix $\Lambda^{ab}$. 
Straightforward but tedious algebra allows us to write down the transformation laws for finite local SU(2) transformations in a concise form
\begin{subequations} \label{SU2Transformation}
  \begin{align}
    n & \to n \\  
   {s}^a & \to {s'}^a = R^{ab} {s}^b \\
    j_\mu & \to j_\mu + \tfrac{1}{c} \tilde{A}_\mu^b s^b ~, \label{jSU2} \\
    J_\mu^a & \to R^{ab} \big[J_\mu^b + \tfrac{1}{c} \tilde{A}_\mu^b n \big] ~, \label{JSU2} \\
    \tau & \to \tau + \tfrac{1}{c} J_\mu^a \tilde{A}_\mu^a + \tfrac{1}{2 c^2} \tilde{A}_\mu^a \tilde{A}_\mu^a ~, \label{SU2t} \\
    \tau^a & \to R^{ab} \big[ \tau^b + \tfrac{1}{c} \tilde{A}_\mu^b j_\mu + \tfrac{1}{2c^2} \tilde{A}_\mu^b \tilde{A}_\mu^c s^c 
    \big] \label{TSU2} \\
    & \phantom{\to} {} - \tfrac{1}{8} R^{ab} (\partial_\mu R^{cb}) (\partial_\mu R^{cd}) s^d
    + \tfrac{1}{4} (\partial_\mu R^{ab}) (\partial_\mu s^b) ~, \nonumber
  \end{align}
\end{subequations}
where we introduced $\tilde{A}_\mu^a = (\partial_\mu \lambda^b) \Lambda^{ba}$, which is the 
non-Abelian gauge vector potential induced by the local SU(2) transformation. 
Note that in the main text, we wrote $\tilde{A}_\mu^a = -\frac{i c}{2} {\rm Tr} \left( \sigma^a U^{\dagger}_{\rm S} \nabla_\mu  U_{\rm S}  \right)$, 
which seems another widely adopted choice in the literature.

Finally, combining the U(1) and SU(2) transformation laws for the density, spin magnetization, and the paramagnetic current and spin current 
-- together with the invariance of $E_{\rm xc}$ -- leads to the transformation properties of the xc potential given in the main text [Eq.\ \eqref{tFields-KS}].

\end{appendix}


\begin{thebibliography}{37}
\bibitem[{not({\natexlab{a}})}]{note1}
\bibinfo{note}{Unless otherwise sated, we use units such that $4\pi\epsilon_0 =
  1$ and $\hbar = e = m = 1$.}
\bibitem[{not({\natexlab{b}})}]{note2}
\bibinfo{note}{However, the corresponding total energies are invariant.}
\end{thebibliography}
\end{document}